\documentclass{article}
\usepackage{amsfonts}
\usepackage{spconf,amsmath,graphicx,epsfig}
\usepackage{subcaption}
\usepackage{mwe}
\usepackage{xcolor}
\usepackage{hyperref}
\usepackage{algorithm}
\usepackage{algorithmic}
\usepackage{adjustbox}
\usepackage{setspace}
\usepackage{dsfont}
\usepackage{bm}
\usepackage{booktabs} %

\DeclareMathOperator*{\argmax}{argmax}

\title{Mitigating Closed-model Adversarial Examples with Bayesian Neural Modeling for Enhanced End-to-End Speech Recognition}
\name{%
\begin{tabular}{@{}c@{}}
Chao-Han Huck Yang$^{1, 2}$\sthanks{Work done as an applied scientist intern at Amazon Alexa AI.} \qquad
Zeeshan Ahmed$^{1}$ \qquad
Yile Gu$^{1}$ \qquad
Joseph Szurley$^{1}$ \\
Roger Ren$^{1}$ \qquad
Linda Liu$^{1}$ \qquad
Andreas Stolcke$^{1}$ \qquad
Ivan Bulyko$^{1}$ \qquad
\end{tabular}}
\address{$^1$Amazon Alexa AI, USA \\
$^2$Georgia Institute of Technology, USA }

\begin{document}
\ninept
\maketitle
\begin{abstract}
In this work, we aim to enhance the system robustness of end-to-end automatic speech recognition (ASR) against adversarially-noisy speech examples. 
We focus on a rigorous and empirical ``closed-model adversarial robustness'' setting (e.g., on-device or cloud applications). The adversarial noise is only generated by closed-model optimization (e.g., evolutionary and zeroth-order estimation) without accessing gradient information of a targeted ASR model directly. We propose an advanced Bayesian neural network (BNN) based adversarial detector, which could model latent distributions against \textbf{adaptive adversarial perturbation} with divergence measurement. We further simulate deployment scenarios of RNN Transducer, Conformer, and wav2vec-2.0 based ASR systems with proposed adversarial detection system. Leveraging the proposed BNN based detection system, we improve detection rate by +2.77 to +5.42\% (relative +3.03 to +6.26\%) and reduce the word error rate by 5.02 to 7.47\% on LibriSpeech datasets compared to the current model enhancement methods against the adversarial speech examples.

\end{abstract}

\section{Introduction}
\label{sec1}
\begin{figure}[ht]
\begin{center}
\vspace{-2mm}
   \includegraphics[width=0.95\linewidth]{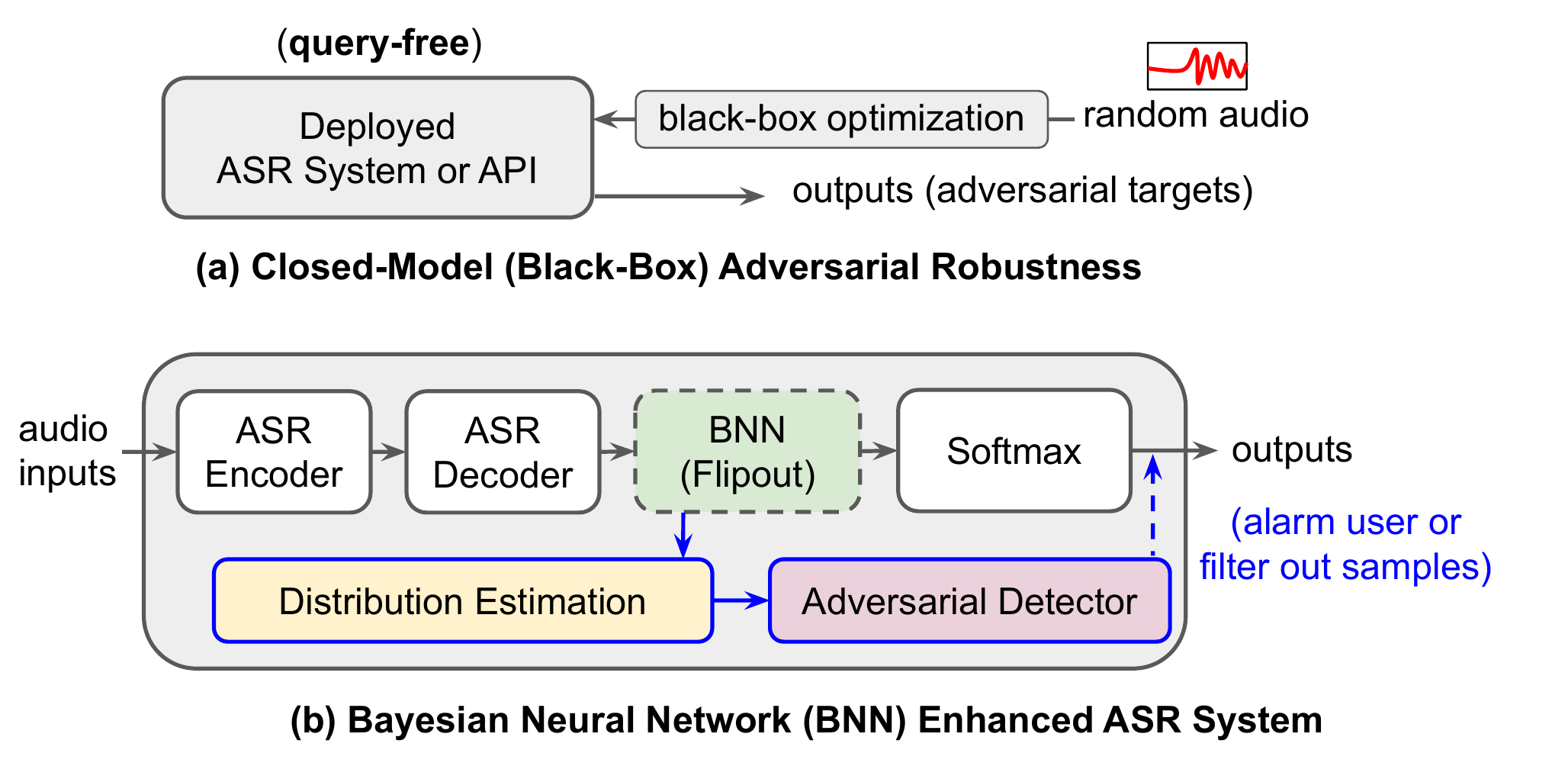}
\end{center}
\vspace{-0.2cm}
   \caption{Illustration of (a)  closed-model adversarial robustness for automatic speech recognition (ASR) from query-based optimization (e.g., on-device or cloud services) and (b) proposed Bayesian neural network (BNN) enhanced model robustness by adversarial detection.   } 
\label{fig:asr_block_diagram}
\end{figure}

End-to-end automatic speech recognition~\cite{lee1990acoustic, graves2006connectionist} (ASR) has many applications in human society, empowering voice-based intelligent control, spoken language understanding~\cite{tur2008calo}, on-device services~\cite{yang2021multi}, and web-based speech interactions~\cite{mendels2015improving}. These high-performance speech applications benefit from neural network-based ASR systems that have highly accurate on-device performance with fixed model parameters. Even when the model parameters of a simulated ASR system are secured by reliable data protection~\cite{yang2021pate}, encryption~\cite{zhang2019encrypted}, and security measures, recent concerns~\cite{yang2020characterizing, wu2021improving} about query-free robustness evaluation highlight the need for designing an ASR system to be robust against noisy samples generated by adversarial optimization. As shown in Fig.~\ref{fig:asr_block_diagram}(a), query-free optimization is applied on random signals to synthesize high-confidence false examples to maliciously manipulate ASR output (e.g., ``open the door''). The aforementioned noise evaluation on a simulated ASR is called closed-model adversarial robustness and could be categorized in a regime of ASR robustness against environmental noise (e.g., acoustic conditions) as a crucial challenge for ASR designs. Table~\ref{table:1:rev} provides an overview of the access constraints for closed-model robustness, which is our focus in this work. 

To improve robustness against adversarial examples, adversarial training-based approaches~\cite{yang2020characterizing, sun2018training} have been widely studied, mainly by applying augmented noisy examples with correct labels to re-train a targeted neural network model to improve its generalization. However, incorporating adversarial training~\cite{sun2018training} into a large-scale neural network is costly and unstable due to its sensitivity to high-dimensional decision boundaries. Furthermore, adversarial training is often considered to not be economical and realistic for online ASR systems, where the model parameters have to be updated frequently, and the augmented noises need to be re-generated each time. 

Recently, adversarial detection mechanisms~\cite{yang2018characterizing, li2021detecting} have emerged as new alternatives to improve system robustness against adversarial examples. Adversarial detection provides low training-cost solutions that are easily incorporated into an existing end-to-end system with an option to control the safety risk by filtering out suspicious inputs. In this work, we take advantage of randomness properties from Bayesian neural networks to design a high-performance adversarial detector empowered by statistical estimation on test samples. As shown in Fig.~\ref{fig:asr_block_diagram}(b), the proposed method is based on an additional variational layer (Flipout~\cite{wen2018flipout}) for distribution estimation that could be easily integrated into end-to-end ASR architectures. Next, we will review existing methods for mitigating adversarial speech examples and demonstrate the novelty of our system design.

\section{Related Work}
\label{section:related-work}

\begin{table}[ht]
\centering
\caption{Type between closed-model ($\mathcal{A}_E$ and $\mathcal{A}_Z$) and open-model ($\hat{\mathcal{A}}$) adversarial robustness evaluation. In this work, we focuses on closed-model settings with access constraints during noise generation.}
\label{tab:1:bvw}
\begin{adjustbox}{width=0.47\textwidth}
\begin{tabular}{|l|c|c|c|c|c|}
\hline
Noise Types & Para. Access & Gradient Info. & Output Access  & Ref. \\ \hline \hline
$\mathcal{A}_E$: Evolutionary & \textbf{No} & No & Yes  & \cite{khare2019adversarial, alzantot2019genattack} \\ \hline
$\mathcal{A}_Z$: Zeroth-order & \textbf{No} & No (estimated) & Yes  & \cite{chen2017zoo, tu2019autozoom} \\ \hline \hline
$\hat{\mathcal{A}}$: Fast Gradient & Yes & Yes & Yes  & \cite{carlini2018audio} \\ \hline
\end{tabular}
\end{adjustbox}
\label{table:1:rev}
\end{table}

\subsection{Adversarial Robustness for Speech Processing}

To study adversarial robustness for ASR, additive noise $\delta$ is applied to the original waveform under an environmental noise ratio = $10$dB reported in the previous evaluation~\cite{yakura2018robust, yang2020characterizing}. The noise is trainable with an objective to degrade model performance or create malicious outputs (e.g., ``open the door''). The adversarial robustness for speech modeling is first studied with a simple open-model setting (white-box), where the $\delta$ is simply generated by gradient information, such as the fast-gradient-sign method (FGSM)~\cite{goodfellow2014explaining} and projected gradient descent (PGD)~\cite{madry2018towards} attacks against the model's decision boundaries. However, open-model adversarial evaluations heavily rely on the hypothesis that the targeted model could be accessed for its gradient information during the noise generation process. This hypothesis is generally \textbf{not} true for real-world speech-based applications in recent discussions and evaluations~\cite{tramer2020adaptive, zelasko2021adversarial, yang2020characterizing}. For example, the end-point device and cloud users cannot extract gradient information from a simulated ASR model directly, where the model is prohibited from external visits and protected by different layers of information security frameworks. 

By contrast, closed-model adversarial evaluations aim to study strict and empirical settings, where the additive environmental noise must be generated without knowing model parameters. Closed-model optimization techniques for the neural model have been used in these studies, such as evolutionary learning~\cite{alzantot2019genattack} and zeroth-order optimization~\cite{tu2019autozoom, chen2017zoo} from sample queries. Multiple-objective evolutionary optimization~\cite{khare2019adversarial} has highlighted closed-model adversarial robustness challenges against popular ASR back-ends, including DeepSpeech~\cite{hannun2014deep} and Kaldi ASR~\cite{povey2011kaldi}. By using only limited query information from input/output pairs, a high acoustic similarity (0.97 to 0.98) is preserved between closed-model generated audio and the original audio, while resulting in large relative word error rate (WER) degradation (up to 980\%). Meanwhile, for improving adversarial robustness, adversarial training is often augmented with correctly labeled noisy samples for training. However, adversarial training requires a long training times and high model complexity. Moreover,  adversarial training is costly for on-device simulation once the ASR model needs updating.  Ideally, parallel model enhancement solutions, including adversarial detection and noise filtering frameworks, would provide tangible, low-complexity, and energy-efficient solutions toward safe and reliable ASR. Next, we will review the existing adversarial detection benchmarks to mitigate the closed-model perturbation for ASR.

\subsection{Distribution Modeling against Closed-Model Perturbation}
To concretely address empirical adversarial robustness, 
\textbf{adaptive noise} is jointly optimized with both targeted and enhancement modules during the adversarial evaluation.  For example, temporal dependency~\cite{yang2018characterizing} (TD) based detection methods have shown the best performance compared to other defensive methods. TD detector randomly divides the inputs sequences into multiple uniform segments and computes the output distribution of whole audio and segmented audio segments for abnormality detection on the adversarial input. More recently, the self-attention mechanism~\cite{yang2020characterizing} has been incorporated into the TD properties and further denoises the adversarial features with multi-scale representation learning. Nevertheless, there is less discussion on how to design an efficient framework to improve the distributional modeling (e.g., abnormal) of the latent representations with the TD properties for the neural ASR model. In this work, we propose a novel design building upon RNN-T, conformer, and wav2vec-2.0 with Bayesian neural networks~\cite{wen2018flipout} to improve distributional modeling. Specifically, incorporating randomness into neural networks has recently been shown to improve smoothness~\cite{li2021detecting, bekasovs2018bayesian} of neural predictors, thus providing stronger robustness guarantees.

\section{Closed-model Adversarial-Robust Neural Speech Recognition}
\label{sec:sec2}

\subsection{Evaluating Closed-model Adversarial Robustness}
\label{sec:adv} 
We first consider an end-to-end classification model $M$, where the classifier is trained with model parameters $\theta$ to produce output prediction (e.g., words or phonemes), $M(\mathbf{x};\theta)=\mathbf{y}$, from an input $\mathbf{x}$ (e.g., acoustic features). The loss function $L_{\theta}(\mathbf{y},\hat{\mathbf{y}})$ is optimized using gradient descent to minimize the prediction error between $\mathbf{y}$ and $\hat{\mathbf{y}}$. In closed-model adversarial robustness, we assume that there is no information about model parameter $\theta$ available to the adversarial noise generator. As shown in Eq.~\ref{eq:1}, untargeted closed-model adversarial noise can only be generated by queries $\mathbf{x},\mathbf{y}$ to maximize the prediction loss with a model outcome of $\mathbf{y}_{adv}$:
\begin{equation}
\mathbf{x}_{a d v} = \mathbf{x} + \delta;~\argmax_\delta \mathbb{E}_{(\mathbf{x}, \mathbf{y}) \sim \mathcal{D}} \left [ L( \mathbf{y}_{adv} , \mathbf{y}) \right ].
\label{eq:1}
\end{equation}
We have fixed the maximum~\cite{alzantot2019genattack} querying number (of input-output pairs $(\mathbf{x},\mathbf{y})$) to 10k in our study, following established closed-model robustness benchmarks. We consider \textbf{adaptive adversarial noise setting} in this work, where the noise is jointly optimized against a simulated ``ASR associated with an adversarial detector'' as a rigorous and firm setting~\cite{tramer2020adaptive, carlini2019evaluating} for empirical adversarial robustness evaluation with a $\delta_{\max}$=0.01 under $l_{\infty}$-norm.

$\mathcal{A}_E$: Multi-objective\textbf{ evolutionary adversarial perturbation}. The method initializes a population of examples  around  the given input example $x$ picking random examples from a uniform distribution defined over the sphere of radius $\delta_{\max}$ centered on the original example. The algorithm computes an adversarial noisy input such that $\left\|\mathbf{x}-\mathbf{x}_{a d v}\right\|_{\infty} \leq \delta_{\max }$. This is achieved by adding random noise in the range $\left(-\delta_{\max }, \delta_{\max }\right)$ to each dimension of the input vector $\mathbf{x}$. We follow the multi-objective optimization method proposed in \cite{khare2019adversarial} to ensure acoustic similarity, computed as the cosine similarly ($\geq$ 0.95) of MFCCs between the noisy and original speech inputs, for the ASR robustness evaluation. Since the work of \cite{khare2019adversarial} does not provide a query number for study, we use the benchmark closed-model algorithm from ~\cite{alzantot2019genattack} to improve the sample efficiency to generate adversarial examples. 

$\mathcal{A}_Z$: Estimated gradient noises with\textbf{ zeroth-order optimization}, where random additive noises are synthesized with estimated gradient information by zeroth-order optimization on open source ASR models. We follow an established zeroth order optimization benchmark proposed by AutoZoom ~\cite{tu2019autozoom} to generated distortion with an adaptive random gradient estimation strategy on the ASR system without accessing the target model parameters. AutoZoom leverages upon jointly optimizing latent embeddings (from an offline encoder) of a target input to generate query-efficient distortion by coordinate-wise gradient estimation based on random vector. We use 
wav2vec2.0~\cite{baevski2020wav2vec} as the offline encoder to extract latent embeddings from inputs speech. The unknown gradient information $\frac{\partial f(\mathbf{x})}{\partial \mathbf{x}_{i}}$ of the $i$-th input speech sample is estimated by:~$
b \cdot \frac{f(\mathbf{x}+\beta \mathbf{u})-f(\mathbf{x})}{\beta} \cdot \mathbf{u} \approx \frac{\partial f(\mathbf{x})}{\partial \mathbf{x}_{i}}$, where $\beta>0$ represents a smoothing parameter (e.g., coordinate-wise estimated gradient); $\mathbf{u}$ is a vector with unit-length. We set the scaling parameter $b = 1$ as in~\cite{duchi2015optimal}. The gradient information was used to compute the standard querying-based closed-model loss function defined in~\cite{chen2017zoo}. We refer to ~\cite{chen2017zoo, tu2019autozoom} for more details on querying-based optimization.

\subsection{Bayesian Neural Network for Adversarial Detection} 
In previous works, BNN and other randomization methods have been used to improve robust classification accuracy, but they were not simulated to improve adversarial detection performance in ASR.

Blundell \emph{et al.}~\cite{blundell2015weight} introduced an efficient algorithm to learn BNN parameters. BNN is subject to model the distribution of the hidden parameters $w$ with the given random variables ($x,y$), instead of estimating the maximum likelihood values $w_{\text{MLE}}$ for the weights. From a Bayesian perspective, each stochastic parameter is now a random variable sampling from a distribution  instead of being a fixed (deterministic) parameter. Given the input $x$ and label $y$, a BNN aims to estimate the posterior over the weights $p(\boldsymbol{w} \mid x, y)$ given the prior $p(\boldsymbol{w})$. The real posterior can be inferred by a parametric distribution $q_{\boldsymbol{\theta}}(\boldsymbol{w})$, where the unknown trainable parameter $\boldsymbol{\theta}$ is estimated by minimizing the KL divergence:~$\mathcal{D}_\mathrm{KL}\left(q_{\boldsymbol{\theta}}(\boldsymbol{w}) \| p(\boldsymbol{w} \mid x, y)\right)$ over $\boldsymbol{\theta}$. $q_{\theta}$ is a factorized Gaussian distribution with neural parameters, where $q_{\theta_i}\left(\boldsymbol{w_i}\right)=\mathcal{N}\left(\boldsymbol{w}_{i};\mu, \sigma^{2}\right)$.

The objective function of the evidence lower bound for training BNNs is reformulated from the expression of KD divergence as shown in (\ref{eq:BBNobjective}), which is a sum of a data-dependent part and a regularization part, for each training data pair $\left(x_{i}, y_{i}\right) \in \boldsymbol{D}$ and $\left(\boldsymbol{w}_{i}\right) \in \boldsymbol{W}$:
\begin{equation}
\underset{\boldsymbol{\theta}}  \argmax \left\{\small \underset{\boldsymbol{W} \sim q_{\theta}}{\mathbb{E}}[\log p(\boldsymbol{D} \mid \boldsymbol{W})]-\mathcal{D}_{\mathrm{KL}}\left(q_{\boldsymbol{\theta}} \| p\right)\right\}
\label{eq:BBNobjective}
\end{equation}
where $\boldsymbol{D}$ represents the data distribution; $\boldsymbol{W}$ represents a random weights distribution. In the first term of objective (\ref{eq:BBNobjective}), the probability of $y_i$ given $x_i$ and weights is the output of the model. This part represents the classification loss. The second term of objective (\ref{eq:BBNobjective}) is trying to minimize the divergence between the prior and the parametric distribution, as a form of regularization.

Based on the theoretical justification and foundations~\cite{li2021detecting} for BNN modeling~\cite{ liu2018adv,tran2019bayesian}, we consider $f(\boldsymbol{x}, \boldsymbol{w})$ as a model with $\boldsymbol{x} \sim \mathcal{C}_{\boldsymbol{x}}$ and $\boldsymbol{w} \sim \mathcal{C}_{\boldsymbol{w}}$, where $\mathcal{C}_{\boldsymbol{w}}$ is any distribution that is symmetric about $\boldsymbol{w}_{0}=\mathbb{E}[\boldsymbol{w}]$, such as $\mathcal{N}\left(\boldsymbol{w}_{0}, \boldsymbol{I}\right) .$ If $\nabla_{\boldsymbol{x}} f(\boldsymbol{x}, \boldsymbol{w})$ can be
approximated by the first-order Taylor expansion at $\boldsymbol{w}_{0}$, we have:
\begin{equation}
\mathcal{W}(f(\boldsymbol{x}+\boldsymbol{\delta}, \boldsymbol{w}), f(\boldsymbol{x}, \boldsymbol{w})) \geq \mathcal{W}\left(f\left(\boldsymbol{x}+\boldsymbol{\delta}, \boldsymbol{w}_{0}\right), f\left(\boldsymbol{x}, \boldsymbol{w}_{0}\right)\right)
\label{eq:5}
\end{equation}
where $\boldsymbol{\delta}$ represents an adversarial perturbation and $\mathcal{W}$ represents a translation-invariant sliced-Wasserstein distance~\cite{kolouri2016sliced} measuring distribution dispersion with one-dimensional closed-form convergence~\cite{kolouri2016sliced}. The inequality performs that parameters associated with randomness will enlarge the distributional differences between normal and adversarial outputs. Thus, we can utilize the distributional differences of the  BNN layer as shown in Fig.~\ref{fig:asr_block_diagram}(b) to detect adversarial examples.  We estimate the  1-Wasserstein distance between the distributions to detect adversarial examples. We show that BNNs can enlarge distributional differences with this distance metric. For a model $f(\boldsymbol{x}, \boldsymbol{w})$ with $\boldsymbol{x} \sim \mathcal{C}_{\boldsymbol{x}}$ and $\boldsymbol{w} \sim \mathcal{C}_{\boldsymbol{w}}$, where $\mathcal{C}_{\boldsymbol{w}}$ is any distribution that satisfies $\boldsymbol{w}$ is symmetric about $\boldsymbol{w}_{0}=\mathbb{E}[\boldsymbol{w}]$, such as $\mathcal{N}\left(\boldsymbol{w}_{0}, \boldsymbol{I}\right)$. We use an efficient stochastic ``Flipout''~\cite{wen2018flipout} method for BNN (as shown in Fig.~\ref{fig:asr_block_diagram}(b)) to sample pseudo-independent posteriors for each input audio, which utilizes a Monte Carlo approximation of the distribution to model neural network parameters. 

\begin{figure}[ht]
\begin{center}
\vspace{-2mm}
   \includegraphics[width=0.85\linewidth]{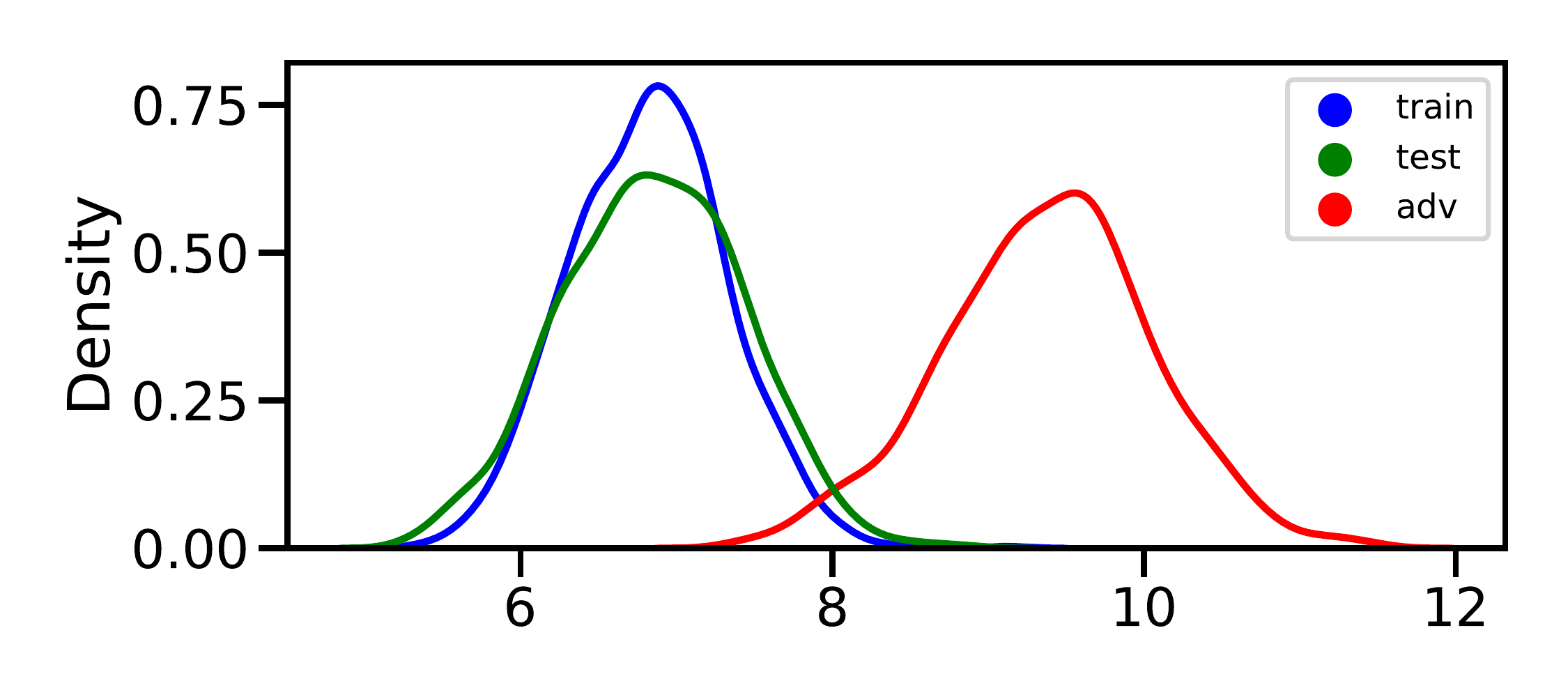}
\end{center}
\vspace{-0.6cm}
   \caption{Standard deviation distributions of hidden layer output of the proposed adversarial detection by Bayesian neural network.  } 
\label{fig:5}
\end{figure}

\textbf{Distribution estimation with BNN:} To interpret the detection behavior of predictions by BNNs, we visualize the distribution of different audio inputs, including the training set (blue), the test set (green), and the adversarial samples (red) during the untargeted noisy evaluation. As shown in Fig.~\ref{fig:5}, the BNN with the aforementioned stochastic modeling is accurate and robust in identifying the adversaries in the LibriSpeech dataset. 

\vspace{-4mm}

\section{Experiments}
\label{sec:exp:content}

\subsection{Training Datasets and ASR Architecture}
\textbf{LibriSpeech dataset:} The LibriSpeech~\cite{panayotov2015librispeech} corpus is a collection of large-scale audiobooks. We use 960 hours of LibriSpeech for our evaluation. For the language modeling results, LibriSpeech-960 also provides the n-gram language models and the corresponding texts taken from Project Gutenberg books, comprising 803M tokens and 977K unique words. We randomly select 2,000 samples from ``dev-clean'' for 
two adversarial robustness evaluation tasks (``targeted words'' or ``non-targeted words'' perturbation) in this study. 
\\

\textbf{ASR architectures for adversarial robustness evaluation:} \linebreak (1) RNN Transformer: We use RNN-T ASR model with five LSTM layers as an encoder. Each LSTM layer has 1024 hidden units. The prediction network in RNN-T comes with two LSTM layers of 1024 units and an embedding layer of 512 units. The one-best sequence is produced by combining the scores of RNN-T with the language model scores. 
We take length-normalized probabilities from the language model (LM) network for its LM scores and determine the coefficient  $\lambda$ using grid search on a development data set; $\lambda$ was fixed at $0.006$ for all reported experiments. 

(2) Conformer ASR: ConformerNet~\cite{gulati2020conformer} has been used in streaming ASR applications. It contains two feed-forward modules sandwiching the multi-headed self-attention module and the convolution module. The first feed-forward module is built from half-step residual weights. The second part of the feed-forward module is connected to a final layer-norm module.

(3) ASR based on wav2vec-2.0 \cite{baevski2020wav2vec}, a pretraining method used to provide waveform-level representations for end-to-end speech recognition. We follow the setup of ``large from scratch'' presented in Baevski \emph{et al.}~\cite{baevski2020wav2vec} to provide a downstream task for wav2vec-2.0 and evaluate its closed-model adversarial robustness.

\subsection{Closed-model Adversarial Robustness Baseline}

\textbf{Task 1 ($\mathcal{T}$1) for untargeted adversarial robustness evaluation:} We first evaluate all the ASR models with (1) the multi-objective evolutionary ($\mathcal{A}_E$)  and (2) zeroth-order ($\mathcal{A}_Z$) estimated gradient information for untargeted noise evaluation. As shown in Eq.~(\ref{eq:1}), the optimization process is based on finding a minimum noise under $l_p$-norm constraints (e.g., environmental noise level in our case) to let the final prediction diverge as far as possible from its original output prediction. The $l_{\infty}$-norm untargeted adversarial robustness evaluation is important, owing to a recent theoretical justification~\cite{weng2018evaluating} of its relationship to an upper bound of model robustness toward other noise (e.g., $l_2$ as Gaussian).

\begin{figure}[ht]
\begin{center}
\vspace{-2mm}
   \includegraphics[width=1\linewidth]{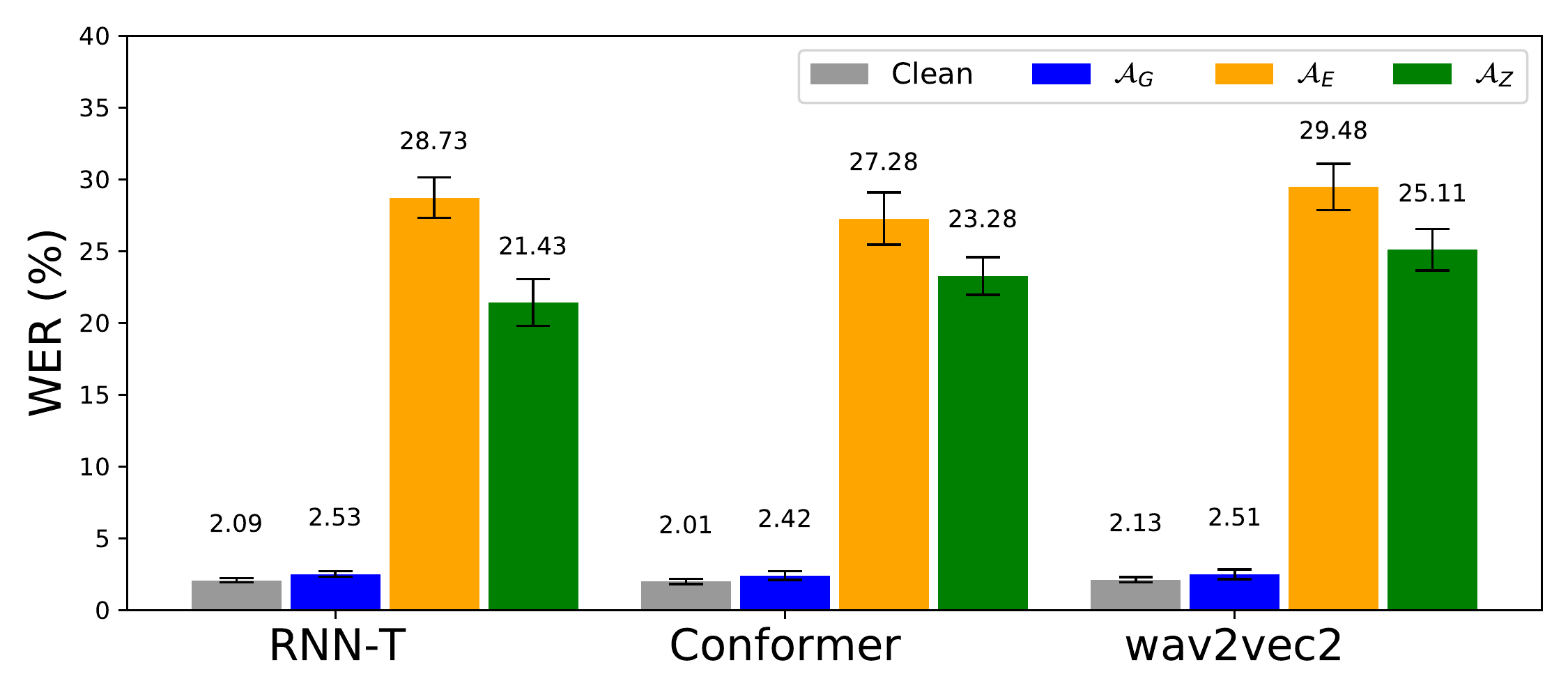}
\end{center}
\vspace{-0.6cm}
   \caption{Task 1 baseline under closed-model adversarial noise.} 
\label{fig:2:baseline}
\end{figure}
\vspace{-2mm}

As shown in Fig.~\ref{fig:2:baseline}, RNN-T, Conformer, and wav2vec-2.0 models show relatively small variances between clean test set and additive Gaussian noise (denoted by $\mathcal{A}_G$) under an SNR of 10dB. However, all the evaluated ASR systems show major performance degradation (increased WER) by evaluating with evolutionary adversarial noise (denoted by $\mathcal{A}_E$) and zeroth-order optimized based adversarial noise (denoted by $\mathcal{A}_Z$) shared the same SNR ratio (10dB) with the additive Gaussian noise setting. While adversarial noises ($\mathcal{A}_E$ and $\mathcal{A}_Z$) causes a severe 23.11\% average WER degradation, Gaussian-based perturbation only leads to a WER increase of 0.38\% to 0.42\%. Notably, the ConformerNet ASR system shows slightly better robustness in terms of WER compared to RNN-T and wav2vec-2.0, which degrades by 25.27\% absolute under $\mathcal{A}_E$ noise and 21.27\% under $\mathcal{A}_Z$ noise. The findings could be due to its scale-free feature learning during the patch-wise processes based on its enhanced convolution architecture, which echos some findings in~\cite{yang2020characterizing}. 

\begin{table}[ht!]
\centering
\caption{Task 1 with detection methods: WER (\%) reduction by filtering out adversarial examples and detection performance.}
\label{tab:t1}
\begin{tabular}{|l|ll|ll|ll|}
\hline
\multicolumn{1}{|c|}{ASR} & \multicolumn{2}{c|}{RNN-T} & \multicolumn{2}{c|}{Conformer} & \multicolumn{2}{c|}{wav2vec-2.0} \\ \hline \hline
Method & TD & BNN & TD & BNN & TD & BNN \\ \hline \hline
WER & -8.41 & \textbf{-13.43} & -9.64 & \textbf{-15.09} & -7.45 & \textbf{14.92} \\ \hline
AUC & 82.31 & \textbf{90.12} & 83.42 & \textbf{92.34} & 80.09 & \textbf{91.09} \\ \hline
FP & 12.23 & \textbf{6.23} & 10.29 & \textbf{5.23} & 14.32 & \textbf{5.78} \\ \hline
FN & 6.43 & \textbf{2.34} & 7.12 & \textbf{2.34} & 8.23 & \textbf{3.23} \\ \hline
\end{tabular}
\end{table}

Next, we integrate an additional adversarial detector into the ASR to detect and filter an \textbf{equally-mixed} of 6k noisy samples (2k for $\mathcal{A}_G$; 2k for $\mathcal{A}_E$; 2k for $\mathcal{A}_Z$). We compare the WER deduction by using a TD detector and a BNN distance-estimated detector in $\mathcal{T}$1 and report the value of WER deduction, area under the curve (AUC), false positive (FP), and false negative (FN) ratios in Table ~\ref{tab:t1}. We observe that Conformer-based ASR is already improved on the four evaluation metrics. The BNN-based detection and filtering system demonstrates superior results compared to the TD detection baseline~\cite{yang2018characterizing}.

\textbf{Task 2 ($\mathcal{T}$2) for targeted adversarial robustness evaluation:} Unlike untargeted perturbation, targeted sentence adversarial robustness evaluation aims to prevent the ASR from making misleading predictions. For example, ''open the door'' is one classical malicious target used in adversarial noise evaluation. To generate noise with a targeted output, the optimization process tries to find a minimal noise signal $\delta$ that minimize prediction loss between noisy prediction $\textbf{y}_{\text{adv}}$ and target $\textbf{y}_{\text{target}}$. The second row of Table~\ref{tab:2} shows that the output predictions of evaluated ASR models are not robust  a mix of noisy Librispeech data. For $\mathcal{T}$2 experiments, we select (1) local smoothing~\cite{yang2018characterizing} (LS), which uses a sliding window of fixed length for local smoothing to reduce the adversarial perturbation; (2)  downsampling~\cite{yang2018characterizing} (DS), where we downsample the original 16\,kHz audio to 8\,kHz, resulting in a audio file with a band-limit, that avoids sacrificing the quality of the recovered audio while reducing the adversarial noise in the reconstruction phase; and (3) TD methods from \cite{yang2018characterizing}. (4) For model defense based on speech enhancement (SE), we select the state-of-the-art self-attention U-Net gated speech enhancement~\cite{yang2020characterizing} against closed-model adversarial noise. As shown in Table ~\ref{tab:2}, the BNN-based detection and filtering outperforms all the existing methods and renders 91.94\% of $\mathcal{A}_E$ and 94.13\% of $\mathcal{A}_T$ adversarial perturbations unsuccessful---a new model robustness benchmark when evaluating with a mix of noisy Librispeech data.

\begin{table}[ht]
\caption{The unsuccessful rate (UR\%) indicating how often adversarial noise fails to manipulate an ASR model with malicious output of ``open the door''. A \textbf{higher} UR value indicates a \textbf{robust} system performance with selected defense algorithms (lines 3 to 5) discussed in Section 4.2. $\mathcal{A}_E$ represents evolutionary noise and $\mathcal{A}_Z$ represents zeroth-order optimized noise as summarized in Table ~\ref{table:1:rev}.}
\label{tab:2}
\vspace{-2mm}
\begin{tabular}{|l|c|c|c|c|}
\hline
     Evaluation         & RNN-T        & Conformer     & wav2vec-2.0         & Avg.        \\ \hline \hline
$\mathcal{A}_E$ noise~\cite{khare2019adversarial}  & 6.76         & 8.79         & 6.39        & 7.31          \\ \hline \hline
$\mathcal{A}_E$ + LS & 10.39 & 10.27 & 10.92 & 10.52 \\ \hline
$\mathcal{A}_E$ + DS & 9.39 & 11.23 & 8.92 & 9.84 \\ \hline
$\mathcal{A}_E$ + TD~\cite{yang2018characterizing} & 88.73 & 86.34 & 84.50 & 86.52 \\ \hline
$\mathcal{A}_E$ + SE~\cite{yang2020characterizing} & 80.99 & 81.42 & 81.34 & 81.25 \\ \hline
$\mathcal{A}_E$ + BNN & \textbf{91.23} & \textbf{93.34} & \textbf{91.25} & \textbf{91.94} \\ \hline \midrule \hline
$\mathcal{A}_Z$ noise~\cite{tu2019autozoom}  & 24.76         & 25.76         & 23.39        & 24.63          \\ \hline \hline
$\mathcal{A}_Z$ + LS & 24.32 & 25.83 & 23.92 & 24.69 \\ \hline
$\mathcal{A}_Z$ + DS & 23.93 & 25.92 & 23.95 & 24.60 \\ \hline
$\mathcal{A}_Z$ + TD~\cite{yang2018characterizing} & 90.87 & 92.34 & 89.12 & 90.77 \\ \hline
$\mathcal{A}_Z$ + SE~\cite{yang2020characterizing} & 91.32 & 93.42 & 89.34 & 91.36\\ \hline
$\mathcal{A}_Z$ + BNN & \textbf{94.53} & \textbf{95.34} & \textbf{92.52} & \textbf{94.13} \\ \hline
\end{tabular}
\end{table}
\vspace{-2mm}

\section{Conclusions}
\label{sec:conclusion}
We have demonstrated a new framework leveraging Bayesian neural networks for latent distribution modeling to improve adversarial robustness for end-to-end speech recognition. 
The newly proposed BNN method attains the best results compared to existing enhancement methods under closed-model adversarial evaluation. The experimental results suggest that recent Conformer and wav2vec-2.0 methods also suffer from the adversarial evaluation challenges for both untargeted and targeted adversarial evaluation. The proposed low-complexity detection method gives us promising new ways to design robust and reliable ASR systems.

\section{Acknowledgment}

The authors thank Ariya Rastrow, Mat Hans and Bj{\"o}rn Hoffmeister from Alexa AI for their valuable comments and discussion. 

\clearpage
\begin{spacing}{0.9}
\footnotesize
\bibliographystyle{IEEEbib}
\bibliography{refs}
\end{spacing}
\end{document}